\documentclass[12pt]{article}

\usepackage{times}
\usepackage{epsf}
\usepackage[numbers,sort&compress]{natbib}
\usepackage{amsmath}
\usepackage{amsfonts}
\usepackage{amssymb}
\usepackage{bm}
\usepackage{bbm}
\usepackage{graphicx}
\usepackage{epsfig}
\usepackage{latexsym}
\usepackage{nicefrac}
\usepackage{xr}

\def\corresponds{{\lower.2ex\hbox{=}}{\rm\kern-.75em^\triangle}}
\def\succsim{\succ\kern-.9em_\sim\kern.3em}
\def\precsim{\prec\kern-1em_\sim\kern.3em}
\def\slantfrac#1#2{\kern1em^{#1}\kern-.3em/\kern-.1em_{#2}}
\def\lfrac#1#2{{}^{#1\!}\kern-.0em/_{#2}}

\def\buildrel#1\under#2{\mathrel{\mathop{\kern0pt #2}\limits_{#1}}}

\newcommand{\slf}[1]{\not{\!#1}}

\begin{document}

\vspace*{0.3cm}

\begin{center}
{\bf \Large Techniques in Analytic Lamb Shift Calculations}\\[2ex]
\normalsize Ulrich D. Jentschura\\
\scriptsize 
{\it \small Max--Planck--Institut f\"ur Kernphysik,
Saupfercheckweg 1, 69117 Heidelberg, Germany\\[0.7ex]
jentschura@mpi-hd.mpg.de}
\end{center}

\begin{abstract}
\noindent
Quantum electrodynamics has been the first theory
to emerge from the ideas of regularization and renormalization,
and the coupling of the fermions to the virtual excitations 
of the electromagnetic field. 
Today, bound-state quantum electrodynamics 
provides us with accurate theoretical predictions
for the transition energies relevant to simple atomic systems,
and steady theoretical progress relies on advances 
in calculational techniques, as well as numerical algorithms.
In this brief review, we discuss one particular aspect 
connected with the recent progress:
the evaluation of relativistic corrections to the one-loop bound-state
self-energy in a hydrogenlike ion of low nuclear charge number,
for excited non-S states, 
up to the order of $\alpha\,(Z\alpha)^6$ in units of the 
electron mass. A few details of calculations formerly
reported in the literature are discussed, and results for 6F, 7F, 6G and 7G
states are given.\\[2ex]
{\bf PACS Nos.:} 31.30.Jv, 12.20.Ds, 11.10.St\\
{\bf Keywords:} Quantum Electrodynamics; Bound States; Atomic Physics.
\end{abstract}

\section{Introduction}	

The bound-state self-energy of an electron in a hydrogenlike atom
has been one of the key theoretical and experimental cornerstones 
in the development of modern field theory. During the 
last decade, the techniques available for 
calculations have dramatically advanced,
and the accuracy of today's predictions is several orders 
of magnitude better than in the early days of quantum
electrodynamics (QED). 
The advances would be impossible without the combined theoretical efforts
toward the description of the hydrogen spectrum, which have involved
generations of physicists. 
The aim of the current brief review is threefold: first, to 
give an overview of some recent advances in 
Lamb shift calculations and self-energy calculations in particular,
second, to describe a few details of recent calculations, 
for higher excited non--S states, which may be useful in an 
independent recalculation of the higher-order binding corrections,
and third, to supplement previously available data for 
higher-order corrections by results for 6F, 7F, 6G and 7G states.

In ``usual'' calculations of virtual loops, we are used 
to the association: ``the number of loops equals the power
of the coupling constant.'' For hydrogenic bound states, the 
situation is different, and the expansion of a bound-state energy, 
including QED effects, is actually a double expansion 
in terms of the QED $\alpha \approx 1/137.036$ and the 
electron-nucleus coupling parameter $Z\alpha$, where $Z$
is the nuclear charge number. 

Specifically,
the leading one-loop energy shifts (due to self-energy and vacuum polarization)
in hydrogenlike systems are of order $\alpha\,(Z\,\alpha)^4 m c^2$,
where $m$ is the electron mass and $c$ is the speed of light.
The complete correction in this order was obtained first in the years 1948 and
1949~\cite{Fe1948,KrLa1949,FrWe1949,Fe1949}.
The leading two-loop energy shifts are of order 
$\alpha^2\,(Z\,\alpha)^4 m c^2$; the paradigm is that
$\alpha$ counts the number of 
loops, whereas $Z\alpha$ is a measure of the relativistic corrections 
that enter the loop corrections and are typical of the bound-state
problem.

When scanning the literature, one should be aware that sometimes, the 
QED energy shifts are expressed in terms of atomic units, 
for which the fundamental energy scale is that of the 
Hartree, which equals $\alpha^2 m c^2$, and 
therefore the leading QED shift
is of order $Z^4 \alpha^3$ in atomic units.

The current paper will be concerned mainly with the calculation
of the correction of order $\alpha\,(Z\,\alpha)^6$, which is effectively
a relativistic correction to the one-loop correction (sic!) due to the 
binding of the electron to the nucleus (hence the name
``binding corrections''). We thus have to consider the 
relativistic atomic physics aspect
of the problem in addition to the usual loop calculation.

The separation of the problem into high- and low-energy
virtual photon contributions has been discussed
(e.g.) in \S 123 of Ref.~\cite{BeLiPi1982}.
In the context of NRQED~\cite{CaLe1986,KiNi1996,NiKi1997,PiSo1998}, 
the general paradigm 
is to map the high-energy effects onto effective operators.
By contrast, the low-energy effects are mapped onto 
the transverse degrees of freedom of the electromagnetic field
and are integrated up to some cutoff scale. 
The systematic expansion of the Wilson coefficients
multiplying the effective operators is rather nontrivial.
In the current context, we start from the fully relativistic 
expression of the self-energy, and do appropriate expansions
for both domains. For the problem at hand, this 
conceptually simpler approach has certain pragmatic 
advantages.

One might well ask why
we would need information on the Lamb shift 
of higher excited states in the first place,
especially because the most accurately measured 
hydrogenic transition involves the 1S and the metastable 
2S level~\cite{NiEtAl2000}.
The answer is that one transition is not enough to determine
fundamental constants like the Rydberg to sufficient
accuracy: one needs at least two transitions.
We would like to cite from Ref.~\cite{UdEtAl1997},
a system of equations that illustrates this fact
in a particularly clear way.
One combines two of the most accurately measured frequencies
($f_{\rm 1S-2S}$~\cite{NiEtAl2000} and a
$f_{\rm 2S-8D}$~\cite{BeEtAl1997} transition),
where we can either use
$8D_{3/2}$ and $8D_{5/2}$ for the $8{\rm D}$ level,
for a determination of the Rydberg.
Defining $e_{\rm D}$ as a dimensionless relativistic 
Dirac energy, we can schematically establish the following
equations,
\begin{subequations}
\label{sys}
\begin{align}
f_{\rm 1S-2S} =&
R_\infty \, c \,
\left\{e_{\rm D}({\rm{2S}}) - e_{\rm D}({\rm{1S}})\right\} +
{\cal L}_{\rm 2S} - {\cal L}_{\rm 1S}\,,
\\[2ex]
f_{\rm 2S-8D} =&
R_\infty \, c\,
\left\{e_{\rm D}({\rm{8D}}) - e_{\rm D}({\rm{2S}})\right\} +
{\cal L}_{\rm 8D} - {\cal L}_{\rm 2S}\,.
\end{align}
\end{subequations}
By $L$, we refer to the Lamb shift [see Eq.~(67) of Ref.~\cite{JePa1996}
for an often-used definition of this effect].
We can eliminate $L_{2S}$ using the combination
\begin{equation}
L_{\rm 2S} = \frac{L_{\rm 1S} + W_{21}}{8}
\end{equation}
where 
\begin{equation}
W_{21} = 8 L_{\rm 2S} - L_{\rm 1S} 
\end{equation}
is independent of the nuclear-size correction 
(proton radius) and can therefore be determined 
theoretically to high accuracy.
We can now solve the system of equations (\ref{sys}) for two
unknowns: $L_{\rm 1S}$ and $R_\infty$,
provided we have theoretical values for the weighted
combination $W_{21}$ and $L_{\rm 8D}$.
The evaluation of an important contribution
to the latter will be one of the issues discussed in the current
brief review.

A generalization of this rather elementary exercise
involves a
least-squares analysis~\cite{JeKoLBMoTa2005}. Roughly speaking, 
one tries to find the ``best fit,'' in the sense of 
least squares, to a set of experimental
results representing the most accurately measured
transitions in hydrogen and deuterium. For the 2002
adjustment~\cite{MoTa2005}, a total of 23 transitions 
in hydrogen and deuterium have been used as
input data. The least-squares analysis includes 
covariances among the theoretical uncertainties. 
One thus obtains theoretical values, including standard 
uncertainties, for the proton radius and the 
Rydberg, and one can make optimal predictions 
for other transition frequencies not used in the 
adjustment of the constants. Some of the predicted frequencies
are currently more accurate~\cite{JeKoLBMoTa2005} than the 
predictions for the anomalous magnetic moment of the 
electron.

There is another aspect which is 
relevant to calculations for higher excited states:
the extrapolation of the data to higher quantum 
numbers, for given manifolds of states.
For example, we can extrapolate to higher 
principal quantum numbers $n$ for given angular momenta
(orbital and spin). 
In an apparently not widely known paper (Ref.~\cite{Po1981}), 
the one-loop Bethe logarithm has been evaluated,
in terms of a (slowly convergent) integral
representation, in the limit of $n \to \infty$.
From this point onwards, the extrapolation to higher 
principal quantum numbers of the 
one-loop nonrelativistic Bethe logarithm, based on a few
known low-$n$ results, becomes an interpolation.
However, this calculation uses a few particular 
properties of the nonrelativistic Bethe logarithm
[order $\alpha(Z\alpha)^4$],
which are not applicable to the 
much more involved correction of order $\alpha\,(Z\alpha)^6$
(relativistic Bethe logarithm).
Nevertheless, it has recently been observed that 
a large number of relativistic effects and loop corrections 
can actually be expressed as an inverse power series 
in the principal quantum number~\cite{Je2003jpa,JeEtAl2003}.
This is not irrelevant because it gives 
confidence that such fits, applied to other 
corrections like the relativistic Bethe logarithm 
provide us with an accurate representation 
of the corrections for large $n$, where the 
complex structure of the wave function
leads to a prohibitively large number of terms in
intermediate steps of the calculation,
which makes an evaluation impossible even if today's 
computer algebra systems~\cite{Wo1988} are used.

This brief review follows two 
investigations~\cite{JeEtAl2003,LBEtAl2003} 
in which previous treatments of the one-loop problem
were extended to higher quantum numbers,
and some general expressions relevant to selected 
manifolds of bound states were presented.
Here, we supplement the results obtained 
previously by explicit numerical calculations
for 6F, 7F, 6G and 7G states. 
Furthermore, in Sec.~\ref{genideas},
we discuss some general ideas relevant to the calculation,
before dwelling on the low-energy part in Sec.~\ref{lowenpart}.
This part, mediated by ultrasoft virtual photons, is 
one of the key elements in the treatment of higher 
excited non-S states. Throughout the paper, we 
restrict the discussion to states with nonvanishing 
orbital angular momentum.
Numerical results are presented in Sec.~\ref{results},
and conclusions are drawn in Sec.~\ref{conclu}.

%
%
\section{General Ideas}
\label{genideas}

We first take the opportunity
to refer the reader to the  two most comprehensive treatments of the one-loop
self-energy problem for a bound state,
which have appeared in the literature
up to now and which we are 
aware of. These are Refs.~\cite{Pa1993} and~\cite{Je2003dipl}.
Here, we briefly mention a few particular aspects of interest
which may be useful in finding a general access to the 
calculations, also with regard to a possible independent 
verification. We fix the units so that $\hbar = c = \epsilon_0 = 1$,
which implies that $e^2 = 4\pi\alpha$. The electron
charge is $e$, not $-e$.

In Feynman gauge, the one-loop bound-state self-energy reads
\begin{equation} 
E = {\rm i} \, e^2 \int_{C_{\rm F}} \frac{d^4 k}{(2 \pi)^4} 
D_{\mu\nu}(k)\,
\left< \bar{\psi} \left| 
\gamma^{\mu} 
\frac{1}{\slf{p} - \slf{k} - m - \gamma^0 V}
\gamma_{\mu}  
\right| \psi \right>
- \left< \bar{\psi} \left| \delta m \right| \psi \right>\,,
\end{equation}
where $D_{\mu\nu}(k)$ is the photon propagator,
and $C_{\rm F}$ is the Feynman contour.
Lorentz covariance is explicitly broken by the 
Coulomb potential $V = -Z\alpha/r$.
The quantity $\delta m$ is the mass counter term,
and $m$ is the electron mass.

We divide the problem of calculation into two contributions,
which correspond to two different photon energy regions,
and two different parts of the photon energy integration
contour $C_{\rm F}$. The photon energy separation 
parameter $\epsilon$
takes the role of an infrared cutoff for the high-energy
part, and it acts as an ultraviolet cutoff for the 
low-energy part. 
The low-energy part may be expressed as
\begin{equation}
E_L = \alpha \sum_{n \geq 4} 
(Z\alpha)^n \, f_n\left( \frac{\epsilon}{(Z\alpha)^2\,m} \right)\,,
\end{equation}
and the high-energy part $E_H$ reads
\begin{equation}
E_H = \alpha \sum_{n \geq 4} 
(Z\alpha)^n \, g_n\left(\frac{ \epsilon}{m} \right)\,.
\end{equation}
The key is to first expand both contributions 
in $Z\alpha$ for small $Z\alpha$, then in $\epsilon$,
for small $\epsilon$. By performing the expansions
in that sequence, we automatically assume
that $(Z\alpha)^2 m \ll \epsilon$.
Counter-intuitively, this is equivalent
to performing an expansion for {\em large}
$\epsilon$, after scaling the powers
of $Z\alpha$ out of the calculation,
as will be demonstrated below in Sec.~\ref{extraction}.
It is also important to realize that 
the actual numerical value of $\epsilon$
is arbitrary. All that matters is the expansion in
systematic expansion in $\epsilon$, including the
logarithmic terms. In Refs.~\cite{Pa1993} and~\cite{Je2003dipl},
it has been stressed that different gauges may be used
for the evaluation of the low- and the high-energy
parts. This gauge ``arbitrariness'' (rather than
``invariance'') holds only if linear 
terms in $\epsilon$ are neglected. This remarkable
fact is a cornerstone in the construction of 
nonrelativistic QED Lagrangians (NRQED), where
most high-energy effective operators are taken from 
calculations carried out in Feynman gauge, but 
the ultrasoft-scale effects (where the virtual photon 
energy is of the order of the atomic binding energy)
are calculated in Coulomb gauge.

The general structure of the one-loop energy shift
$E$, for non-S states, is
\begin{equation}
E = E_L + E_H = 
\frac{\alpha}{\pi} \,
\frac{(Z\alpha)^4 \, m_{\rm e} \,c^2}{n^3} \,
F(Z\alpha)\,.
\end{equation}
The dimensionless function $F(Z\alpha)$ depends on the 
bound-state quantum numbers and has the expansion
(again, for non-S states)
\begin{equation}
F(Z\alpha) = A_{40} + (Z \alpha)^2 \, \left\{ A_{61} \,
\ln\left[(Z \alpha)^{-2}\right] + A_{60}\right\}\,.
\end{equation}
Here, we ignore higher-order terms irrelevant for the current
$\alpha\,(Z\alpha)^6$-calculation.
The indices of the constant coefficients $A_{XY}$ 
correspond to the power $X$ of $Z\alpha$ and 
the power $Y$ of the logarithm $\ln[(Z\alpha)^{-2}]$.
The function $F(Z\alpha)$ is obtained as the sum
\begin{equation}
\label{F}
F(Z\alpha) = F_H(Z\alpha, \epsilon) 
+ F_L(Z\alpha, \epsilon) \,,
\end{equation}
where $F_H$ corresponds to $E_H$
and $F_L$ corresponds to $E_L$.
The sum $F(Z\alpha)$ is independent of the cutoff parameter
$\epsilon$.

Let us consider the concrete case
of the $8{\rm D}_{3/2}$ state.
The high-energy part reads
\begin{equation}
F_{\rm H}(8{\rm D}_{3/2}) =
- \frac{1}{20} +
(Z\alpha)^2 \, \left[-\frac{20893}{2419200}
- \frac{31}{2520}\,\frac{m}{\epsilon}
- \frac{31}{2520} \, \ln\left(\frac{2\epsilon}{m}\right) \right]\,.
\end{equation}
The calculation entails techniques familiar from high-energy
physics and is described in detail in Ref.~\cite{Je2003dipl}.
The term $-1/20$ is a consequence of the 
anomalous magnetic moment of the electron as described in detail
in Chap.~7 of Ref.~\cite{ItZu1980}. The low-energy part is 
obtained largely by atomic-physics calculational
techniques and reads
\begin{align}
F_{\rm L}(8{\rm D}_{3/2}, Z\alpha, \epsilon)
=& - \frac43 \ln k_0 (8 {\rm D}) 
\nonumber\\[2ex]
& + (Z\alpha)^2 \left[0.024886
+ \frac{31}{2520}\,\frac{m}{\epsilon}
+ \frac{31}{2520} \ln\left(\frac{\epsilon}{(Z\alpha)^2 m}\right)
\right],
\end{align}
where $\ln k_0$ is the familiar nonrelativistic Bethe 
logarithm. The sum is
\begin{align}
\label{concrete8D32}
F(8{\rm D}_{3/2}, Z\alpha,\epsilon) =&
F_{\rm H}(8{\rm D}_{3/2}, Z\alpha) +
F_{\rm L}(8{\rm D}_{3/2}, Z\alpha)
\nonumber \\[2ex]
=& - \frac{1}{20} - \frac43\,\ln k_0 (8 {\rm D})
+ (Z\alpha)^2 \, \left[
\frac{31}{2520} \, \ln\left[(Z\alpha)^{-2}\right]
+ 0.007723 \right]
\end{align}
and thus free of $\epsilon$, as it should be.
A comparison of Eqs.~(\ref{F}) and (\ref{concrete8D32}) 
reveals that
\begin{equation}
A_{60}(8{\rm D}_{3/2}) = + 0.007723 \,.
\end{equation}
The generalization of the term 
$-1/20 - (4/3)\,\ln k_0 (8 {\rm D})$
to an arbitrary hydrogenic state is discussed in~\ref{massren}.
Model examples illustrating the cancellation of the 
cutoff parameter $\epsilon$ can be found in Appendix~A 
of Ref.~\cite{JePa2002} and in Section 2 of 
Ref.~\cite{JeKePa2002}.

%
%
\section{Low--Energy Part}
\label{lowenpart}

%
%
\subsection{Orientation}
\label{orient}

In this brief review, we will focus on the 
low-energy part (``ultrasoft'' scale, $\omega \sim (Z\alpha)^2\,m$).
According to Eq.~(3.109) of Ref.~\cite{Je2003dipl},
this term may be written as
\begin{equation} 
\label{EL}
E_L = - \frac{e^2}{2}
\int_0^\epsilon \frac{d k\,k}{(2\pi)^3}
\int {\rm d}\Omega_{\bm k}
\left( \delta^{ij} - \frac{k^i \, k^j}{k^2} \right)
\left< \psi^+ \left| 
\alpha^{i} {\rm e}^{{\rm i} {\vec k}\cdot {\vec r}}
\frac{1}{H_{\rm D} - E_{\rm D} + k}
\alpha^{i} {\rm e}^{ -{\rm i} {\vec k}\cdot {\vec r}}
\right| \psi \right>,
\end{equation}
where the fully relativistic Dirac wave functions
$\psi$ is used, and the $\alpha^i = \gamma^0 \gamma^i$
matrices are a noncovariant form of the Dirac $\gamma$ matrices.
The quantity $k = |\vec k|$ is virtual photon energy.
Note that one should actually supplement a 
principal value prescription in this formula,
because the lower-lying states generate
poles in the integration region $k \in (0,\epsilon)$
with $\epsilon \gg (Z\alpha)^2 m$.
The ensuing problem of the accurate definition 
of a resonance eigenvalue has been discussed in 
Ref.~\cite{JeEvKePa2002}, in the context of 
two-loop corrections.

It is well known that the Dirac Hamiltonian $H_{\rm D}$ 
in Eq.~(\ref{EL}) can be transformed to the Schr\"{o}dinger 
Hamiltonian plus various relativistic corrections,
using the Foldy--Wouthuysen transformation $U$,
which leads to a transformed Hamiltonian
$U H_{\rm D} U^{-1}$.
It is this transformation which is the key to the 
successful identification of all terms which contribute
to the low-energy contribution in a given order of the 
$Z\alpha$ expansion. Specifically, 
one may apply that same transformation $U$ to
the current operator~\cite{JePa1996},
\begin{equation}
j^j = \alpha^j \exp({\rm i} \, \vec{k} \cdot \vec{r})\,.
\end{equation}
The result is
\begin{equation}
U j^j U^{-1} = \frac{p^j}{m} + \delta j^j,
\end{equation}
where
\begin{eqnarray}
\label{alphaitransformed}
\delta j^j & \sim &
\frac{p^j}{m} \,
\left(1 + {\rm i}\, \left( {\vec k} \, \cdot {\vec r} \right) -
\frac{1}{2} \left( {\vec k} \, \cdot \, {\vec r} \right)^2 \right) \\
& & - \frac{1}{2 \, m^3} p^j \vec{p}^{\,2} -
\frac{1}{2 \, m^2} \, \frac{Z\,\alpha}{r^3} \, 
  \left( {\vec r} \times \vec{\sigma} \right)^j \nonumber \\
& & + \frac{1}{2 \, m} \left( {\vec k} \, \cdot \, {\vec r} \right)
 \left( {\vec k} \times \vec{\sigma} \right)^j \nonumber 
- \frac{\rm i}{2 \, m}
   \left( {\vec k} \times \vec{\sigma} \right)^j. \nonumber
\end{eqnarray}
Here, we have dropped terms which couple lower
and upper components of the Foldy-Wouthuysen transformation
and lead to vanishing energy shifts within the context
of the $\epsilon$-expansion.
The term proportional to
$ {\vec k} \times \vec{\sigma}$ entails a spin flip,
but the leading-order current $p^j/m$ necessitates a 
change in the angular momentum of the electron. 
Because of angular momentum selection rules,
we may thus neglect this term in the 
calculation.

Alternatively, one might have obtained the current 
(\ref{alphaitransformed}) by considering a Foldy--Wouthuysen 
transformation of a Dirac Hamiltonian 
with a vector potential $\vec A$ included, i.e.
a transformation of $\vec{\alpha}\cdot 
(\vec{p} - e\vec{A})+ \beta\,m + V$ instead 
of $\vec{\alpha}\cdot \vec{p} + \beta\,m + V$. 
The effective current operator then is the 
term that multiplies $\vec A$ in the transformed Hamiltonian.
That latter approach is used, e.g., in Ref.~\cite{Pa2004}.
The term
\begin{equation}
\frac{p^j}{m} \,
\left(1 + {\rm i}\, \left( {\vec k} \, \cdot {\vec r} \right) -
\frac{1}{2} \left( {\vec k} \, \cdot \, {\vec r} \right)^2 \right) 
\end{equation}
leads to the so-called quadrupole correction.
The other current corrections in Eq.~(\ref{alphaitransformed})
are considered separately and identifies as 
``relativistic corrections to the current'' (see 
Refs.~\cite{JePa1996,JeSoMo1997}).
Of course, these corrections
have to be supplemented by 
relativistic corrections to the 
Hamiltonian and to the bound-state energy, 
as well as corrections to the wave function.
These corrections are well defined and identify
all terms relevant in the order 
$\alpha\,(Z\alpha)^6\,m$.

%
%
\subsection{Dimensionless energy parameter}
\label{dimless}

One of the advantages of the Foldy--Wouthuysen 
transformation is that we may carry out all further
calculations using the nonrelativistic 
form of the bound-state propagator.
There exists a closed-form 
Sturmian decomposition for
all angular components~\cite{SwDr1991a,SwDr1991b,SwDr1991c},
\begin{equation}
\label{green1}
g_l(r_1, r_2, \nu) = \frac{4 m}{a_0 \nu} 
\left( \frac{2 r_1}{a_0 \nu} \right)^l \, 
\left( \frac{2 r_1}{a_0 \nu} \right)^l \,
e^{- (r_1 + r_2)/(a_0 \nu) } 
\sum_{k=0}^{\infty} 
\frac{L_k^{2 l + 1}\left(\frac{2 r_1}{a_0 \nu} \right) \, 
L_k^{2 l + 1}\left(\frac{2 r_2}{a_0 \nu} \right)}
{(k+1)_{2l+1} \, (l + 1 + k - \nu)}.
\end{equation}
Here, $a = 1 / (Z\alpha m)$
is the Bohr radius, and $(k)_c$ is the Pochhammer symbol.  
The symbols $L_k^{2l+1}$ denote the associated Laguerre polynomials. 
The radial integrals can usually be evaluated 
using standard techniques, which leaves the sum over $k$
as a final problem in the calculation (note that $k$ in this 
context is to be differentiated from the virtual 
photon energy which is also denoted by $k$ in many calculations
in this field of research).
In~\ref{hypergeom}, we discuss
some important properties of typical hypergeometric function
encountered which result from an evaluation the sums.
Convergence acceleration techniques are useful for 
accurate calculations of relativistic effects,
because the $k$-sum is typically slowly convergent for high virtual
photon energies.

The nonrelativistic Schr\"{o}dinger--Coulomb Green
function for the hydrogen atom~\cite{SwDr1991b} reads
\begin{equation}
\label{green2}
\left< \bm{r}_1 \left| \frac{1}{H-z} \right| \bm{r}_2 \right> =
\sum\limits_{lm} g_l(r_1, r_2;\nu)\,
Y_{lm}(\theta_1,\varphi_1)\,Y^*_{lm}(\theta_2,\varphi_2)\,.
\end{equation}
Here, $z = E_{\rm S} - k$ is an energy parameter that 
involves the Schr\"{o}dinger energy $E_{\rm S}$ of the 
reference state. When the photon energy $k$ assumes
values in the range $(0, \infty)$, the argument $z$
decreases monotonically from $E_{\rm S} < 0$ to
$-\infty$. The basic idea is now to set
$t = \sqrt{E_{\rm S}/z}$, so that $t$ runs from a value $t=1$
for $k=0$ to a value $t = 0$ for $k =  \infty$.
As a function of $Z\alpha$ and the principal
quantum number $n$, the quantity
$t$ parameterizes the energy argument $z$ as
\begin{subequations}
\label{tparam}
\begin{eqnarray}
z &\equiv& z(t) = -\frac{(Z\alpha)^2 \, m}{2 \, n^2 \, t^2}\,,\\[2ex]
t &\equiv& t(z) = \frac{Z\alpha}{n} \,
\sqrt{-\frac{m}{2z}}\,.
\end{eqnarray}
\end{subequations}
Equation (3.126) of Ref.~\cite{Je2003dipl}
specializes this transformation to the
case $n=2$. In Refs.~\cite{SwDr1991a,SwDr1991b,SwDr1991c}, 
we encounter the notation
\begin{equation}
\nu = n \, t = (Z\alpha) \, \sqrt{-\frac{m}{2z}}\,,
\end{equation}
which enters into Eqs.~(\ref{green1}) and~(\ref{green2}).
An expansion for large $k$ then corresponds to an 
expansion for small $t$ (see~\ref{hypergeom}).

%
%
\subsection{Extraction of a finite part}
\label{extraction}

In many cases, it is necessary, within bound-state calculations,
to extract a nonlogarithmic, constant term from an
integral that contains a variety of divergent contributions 
in the ultraviolet domain. Here, we discuss 
possible algorithms which can be used for such calculations, 
and which may eventually be useful for an independent verification
of the results reported here. For example, the finite parts
which lead to the relativistic Bethe logarithms might be 
alternatively extracted using a discretized Schr\"{o}dinger--Coulomb
propagator on a lattice~\cite{SaOe1989}. In this case,
it is necessary to have a means for extracting nonlogarithmic 
terms numerically rather than analytically. We believe that an 
independent verification of the nonlogarithmic terms, 
using purely numerical methods, could be a very 
worthful cross-check of the analytic approach (up to the last
stage of the calculation where the integrals are evaluated
numerically) which was used 
in previous evaluations.

In order to approach this problem, 
we first consider a model calculation inspired by the 
structure of the nonrelativistic integrand that leads 
to the Bethe logarithm in Eq.~(\ref{bethelog}) below.
It consists in the evaluation of the integral
\begin{align}
\label{intI}
I =& \int_0^\Lambda {\rm d}k \, k \,\frac{(Z\alpha)^2}{k + h (Z\alpha)^2}
\nonumber\\[2ex]
=& (Z\alpha)^2 \Lambda -
(Z\alpha)^4 \, h \, \ln\left( 
\frac{\Lambda + (Z\alpha)^2 h}{(Z\alpha)^2 h} \right)
\end{align}
and in the extraction of the logarithmic and nonlogarithmic 
terms ($\epsilon$ prescription), or in the calculation of 
just the nonlogarithmic term (numerical evaluation 
of generalized Bethe logarithms).

The first prescription would consist in 
expanding the integral for large $\Lambda$, and dropping all
linear, quadratic etc.~terms for large $\Lambda$.
In this way, we obtain the result
\begin{equation}
\label{resforI1}
I \sim -(Z\alpha)^4 \, h \, \ln\left( 
\frac{\Lambda}{(Z\alpha)^2 h} \right)\,.
\end{equation}
Let us now consider a variation of the first prescription.
We go to atomic units, i.e.~scale all powers of $Z\alpha$
out of the integrand via the transformation $k \to (Z\alpha)^2 k$. 
We then define $\lambda = \Lambda/(Z\alpha)^2$ and obtain
\begin{align}
\label{resforI2}
I =& (Z\alpha)^4 
\int_0^\lambda {\rm d}k \, k \,\frac{1}{k + h}
\nonumber\\[2ex]
=& (Z\alpha)^4 \, h \, \left[ \lambda -
\ln\left( \frac{h + \lambda}{h} \right) \right]
\nonumber\\[2ex]
\sim& -(Z\alpha)^4 \, h \, 
\left[ \ln\left( \frac{\lambda}{h} \right) \right]\,,
\end{align}
where the last form is obtained after dropping the 
linear and (possibly further) 
quadratic terms; the final result is 
in full agreement with Eq.~(\ref{resforI1}).

The second prescription is based on the 
identification $\Lambda \to \epsilon$. We first expand 
the result in Eq.~(\ref{intI}) in powers of
$Z\alpha$, which entails the replacement
\begin{equation}
\frac{\epsilon + (Z\alpha)^2 h}{(Z\alpha)^2 h} \to
\frac{\epsilon}{(Z\alpha)^2 h} 
\end{equation}
in the argument of the logarithm in Eq.~(\ref{intI}).
We then expand in $\epsilon$ for small $\epsilon$, dropping
the linear terms. We thus obtain the result
\begin{equation}
\label{resforI3}
I \sim -(Z\alpha)^4 \, h \, \ln\left( 
\frac{\epsilon}{(Z\alpha)^2 h} \right)\,,
\end{equation}
which is equivalent to the result in 
Eq.~(\ref{resforI1}) upon the 
identification $\Lambda \to \epsilon$.
This illustrates that the expansion in small $\epsilon$ after
the expansion in $Z\alpha$ is actually an expansion
for $\epsilon \gg (Z\alpha)^2 m$, after dropping the linear terms.
In particular, all of the above prescriptions lead to the 
result 
\begin{equation}
I \sim (Z\alpha)^4 \, h \, \ln\left( (Z\alpha)^2 h \right)\,,
\end{equation}
for the nonlogarithmic term.

Let us now suppose that we can evaluate a general function $f(x)$
only numerically, but that we know its expansion for large
$x$,
\begin{equation}
\label{deff}
f(x) = \sum_{n = -1,-\nicefrac{1}{2},0,\nicefrac{1}{2},\dots,m} a_n \, x^n + 
{\cal O}(x^{-3/2}), \qquad x \to \infty\,,
\end{equation}
and we wish to evaluate the nonlogarithmic term $N$ that is
generated by the integral 
\begin{equation}
\int_0^\Lambda {\rm d}x \,f(x) \sim N
\end{equation}
for large $\Lambda$. In the sense of Eq.~(\ref{intI}),
we would have $f(x) = x (Z\alpha)^2/[x + h (Z\alpha)^2]$ and 
and $N =  (Z\alpha)^4 \, h \, \ln\left( (Z\alpha)^2 h \right)$.
For the more general case [see Eq.~(\ref{deff})], we define
\begin{equation}
d(x) = \sum_{n = -1,-\nicefrac{1}{2},0,\nicefrac{1}{2},\dots,m} a_n \, x^n
\end{equation}
and
\begin{equation}
D(x) = a_{-1} \, \ln(x) +
\sum_{n = -\nicefrac{1}{2},0,\nicefrac{1}{2},\dots,m} 
\frac{a_n \, x^{n+1}}{n+1}\,.
\end{equation}
For arbitrary $M$, the nonlogarithmic term $N$
may then be extracted according to 
\begin{align}
N =& {\cal I}_1 + {\cal I}_2 + {\cal I}_3\,,
\end{align}
with
\begin{align}
{\cal I}_1 =& \int_0^M {\rm d} x \, f(x)\,,
\nonumber\\[2ex]
{\cal I}_2 =& \int_M^\infty {\rm d} x \, [f(x) - d(x)]\,,
\nonumber\\[2ex]
{\cal I}_3 =& -D(M)\,.
\end{align}
The sign of the ${\cal I}_3$-term
is determined by the necessity to subtract the integral $D(x)$
of the subtraction term $d(x)$ at the lower limit
of integration $x=M$. This consideration
effectively results in three minus signs.
Analogous considerations have recently been used in 
Refs.~\cite{PaJe2003,Je2004b60}.

%
%
\section{Numerical Results}
\label{results}

For the states under investigation, the $A_{60}$ coefficients
are listed in Tables 1 and 2.
For the $5F_{5/2}$ state, the result 
had previously been recorded as $0.002~403~158$,
and for $5F_{7/2}$, a value of $0.008~087~020$
had been indicated (see Ref.~\cite{LBEtAl2003}).
The correction of this result, in the last decimal, 
is beyond current and projected levels of experimental
accuracy. For the current brief review, we re-evaluate 
many of the integrals leading to the $A_{60}$ coefficients
with an enhanced number of integration nodes.
The two entries in question
change by more than the previously indicated 
numerical accuracy. Results for 6F, 7F, 6G, and 7G as reported 
in Tables 1 and 2 are obtained here.
In Ref.~\cite{JeEtAl2003}, we already corrected 
a computational error for $3{\rm P}_{1/2}$ as previously reported in
Eq.~(96) of Ref.~\cite{JeSoMo1997}, where a value of $-1.14768(1)$ 
had been given. As in previous
calculations (see Refs.~\cite{JePa1996,JeSoMo1997}), certain remaining
one-dimensional integrals
involving (partial derivatives of) hypergeometric functions could only be
evaluated numerically. For $n = 8$, we recall the results~\cite{JeEtAl2003} 
$A_{60}(8{\rm D}_{3/2}) = 0.007~723~850$ and
$A_{60}(8{\rm D}_{5/2}) = 0.034~607~492$.

\begin{table}[ht]
\begin{center}
\begin{minipage}{14cm}
\begin{center}
\caption{A table of $A_{60}$ coefficients for
higher excited atomic states with positive Dirac angular
quantum number $\kappa$ (i.e.,~$j = l-1/2$).
All decimal figures shown are
significant.}
\begin{tabular}{crrrr}
\hline
\hline
$n$ & $A_{60}(n{\rm P}_{1/2})$ & $A_{60}(n{\rm D}_{3/2})$ & 
$A_{60}(n{\rm F}_{5/2})$ & $A_{60}(n{\rm G}_{7/2})$ \\
\hline
2 & $-0.998~904~402$ & -- & -- & -- \\
3 & $-1.148~189~956$ & $0.005~551~573$ & -- & -- \\
4 & $-1.195~688~142$ & $0.005~585~985$ & $0.002~326~988$ & -- \\
5 & $-1.216~224~512$ & $0.006~152~175$ & $0.002~403~151$ & $0.000~814~415$ \\
6 & $-1.226~702~391$ & $0.006~749~745$ & $0.002~531~636$ & $0.000~827~468$ \\
7 & $-1.232~715~957$ & $0.007~277~403$ & $0.002~661~311$ & $0.000~857~346$ \\
\hline
\hline
\end{tabular}
\end{center}
\end{minipage}
\end{center}
\end{table}

\begin{table}[ht]
\begin{center}
\begin{minipage}{14cm}
\begin{center}
\caption{Analog of Table~1 for states with negative $\kappa$.}
\begin{tabular}{@{}crrrr@{}}
\hline
\hline
$n$ & $A_{60}(n{\rm P}_{3/2})$ & $A_{60}(n{\rm D}_{5/2})$ & 
$A_{60}(n{\rm F}_{7/2})$ & $A_{60}(n{\rm G}_{9/2})$ \\
\hline
2 & $-0.503~373~465$ & -- & -- & -- \\
3 & $-0.597~569~388$ & $0.027~609~989$ & -- & -- \\
4 & $-0.630~945~795$ & $0.031~411~862$ & $0.007~074~961$ & -- \\
5 & $-0.647~013~508$ & $0.033~077~570$ & $0.008~087~015$ & $0.002~412~929$ \\
6 & $-0.656~154~893$ & $0.033~908~493$ & $0.008~610~109$ & $0.002~748~250$\\
7 & $-0.662~027~568$ & $0.034~355~926$ & $0.008~906~989$ & $0.002~941~334$ \\
\hline
\hline
\end{tabular}
\end{center}
\end{minipage}
\end{center}
\end{table}

%
%
\section{Conclusions}
\label{conclu}

The challenges 
of bound-state quantum electrodynamic calculations
are associated to the accuracy
of the experimental verifications, to the significance of the
theory for the determination of the fundamental constants, and to the
conceptual complexity of the calculations
which is derived from the apparent simplicity of the
physical systems under study. The latter aspect
is developed to full extent only if an accurate understanding
is required in higher orders of perturbation theory.

Let us briefly discuss possible extensions of this work
in addition to the independent verification using a
more numerically inspired approach, as outlined in 
Sec.~\ref{extraction}.
In Refs.~\cite{JePa2002,Pa1999}, calculations of the fine-structure
splitting for P states are described which rely on a 
form-factor approach; this would be equivalent to 
using an effective operator for the high-energy 
part. Usually, NRQED-inspired calculations involve a high-energy
part, which takes care of the contribution of the
high-energy virtual photons, and which is given by effective
operators, and a low-energy part, which is given
by photons whose energy is of the order of the electron
binding energy. The latter integration region is often 
referred to as the ``ultrasoft scale'' in the literature
(see, e.g.,
Refs.~\cite{KiNi1996,NiKi1997,PiSo1998,Pi2002pra,Pi2002prd}).
The two scales (sometimes three, if one makes
an additional distinction with regard to electron momenta)
require a completely separate treatment and
cannot be calculated on the same footing.
Thus, the introduction of scale-separation parameters
is required. These cancel at the end of the calculation.
For the high-energy effective operators, this scale-separation
parameter takes the role of an infrared cutoff, whereas
for the low-energy contributions, the scale-separation
parameter gives a natural scale for the failure of the
nonrelativistic (``ultrasoft'') approximation to the virtual
photons, i.e.~it acts as an ultraviolet cutoff.
This property is characteristic of QED bound-state calculations
and is a feature that adds a certain twist to the analysis
which is not present in usual quantum-field theoretic
loop calculations.

\appendix

%
%
\section{Mass renormalization and the leading-order result}
\label{massren}

In the first articles on the
self-energy, the concepts of mass renormalization and covariant
integration were developed and applied for the first time
to the calculation of observable physical effects. The leading-order
energy shift, of order $\alpha(Z\alpha)^4 m\,c^2$,
is the sum of a Dirac $F_1$ form-factor contribution,
an anomalous magnetic moment term ($F_2$ electron form factor),
a vacuum-polarization term, and an effect mediated
by low-energy virtual photons. The latter contribution can be interpreted
as an ``average virtual excitation energy'', 
as a ``logarithmic sum'' or, as it is most widely 
called, a ``Bethe logarithm''
$\ln k_0$. One may therefore point out that the leading 
order-$\alpha(Z\alpha)^4$-effect
summarizes already the core of most properties of the electron
mediated by the virtual interactions with the quantum fields.
The self-energy of a bound electron is actually the difference
of the self energies of a bound and a free electron, the latter
being attributable to its mass, wherefore it can be
reabsorbed into a redefined physical parameter entering
the Lagrangian of QED.

Yet at the same time, it is important 
to remember that different energy scales enter the
problem: (a) the atomic binding energy and
(b) the relativistic electron rest mass energy scale.
The separation is already explained in 
\S 123 of Ref.~\cite{BeLiPi1982},
and its extension to higher orders in the $Z\alpha$ expansion
is discussed in this brief review.

The generalization 
of the $\alpha(Z\alpha)^4 m$-part of the 
result in Eq.~(\ref{concrete8D32}) 
to an arbitrary non-S state, reads
\begin{equation}
E \sim 
\frac{\alpha}{\pi} \, \frac{(Z\alpha)^4 \, m}{n^3} \,
\left( -\frac{1}{2\,\kappa\,(2 l + 1)} - \frac43 \, \ln k_0(n,l) \right) \,.
\end{equation}
The $A_{40}$-coefficient for a non-S state
thus reads (see, e.g., Refs.~\cite{SaYe1990} and~\cite{MoTa2000})
\begin{equation}
\label{A40gen} 
A_{40} = -\frac{1}{2\kappa\,(2 l + 1)} - \frac{4}{3}\,\ln k_0(nl),
\end{equation}
where $\kappa = 2 (l-j) (j+1/2)$, with 
the usual meaning for the bound-state angular momentum
quantum numbers.
The Bethe logarithm $\ln k_0(nl)$ is an inherently nonrelativistic
quantity, whose expression in natural units reads
\begin{equation}
\label{bethelog}
\ln k_0(nl) = \frac{n^3}{2 (Z\alpha)^4\,m} 
\left< \phi \left| \frac{p^i}{m} \,
\left( H_{\rm S} - E_n \right) \, 
\ln \left[ \frac{2 \left| H_{\rm S} - E_n \right|}{(Z\alpha)^2\,m} \right] \,
\frac{p^i}{m} \right| \phi \right>\,. 
\end{equation}
Here, $\phi$ is the nonrelativistic
(Schr\"{o}dinger) form of the bound-state wave function.

It took 50 years~\cite{Be1947,Fe1948,KrLa1949,FrWe1949,Fe1949} 
to advance our understanding 
from the order $\alpha (Z\alpha)^4 m_{\rm e}c^2$
to the order $\alpha (Z\alpha)^6 m_{\rm e}c^2$.
For a successful calculation
of the higher-order relativistic corrections, 
it has been essential to 
master the much more involved analytic, covariant integrations
in the high-energy part,
and to advance our understanding
of relativistic and retardation corrections 
to the current. Also, the possibility of handling very involved analytic 
intermediate expressions by computer algebra
has become an essential ingredient of the modern 
calculations.

Finally, let us remark 
that for low-lying states, it is possible to evaluate the one-loop
effect numerically to very high 
accuracy~\cite{JeMoSo1999,JeMoSo2001pra,JeMo2004pra},
even for low nuclear charge numbers. 
For very highly excited states, however, first 
exploratory work has revealed that the 
numerical difficulties associated with the 
renormalization are still rather prohibitive 
in the context of an accurate numerical
evaluation, because it entails a loss of more than twelve
decimals for higher excited states. 
In addition, the wave functions become much more complex,
and this inhibits a fast convergence of the 
numerical integrations,
leading to the temporary conclusion
that analytic calculations are still preferable
for higher excited states, in the domain of 
low nuclear charge numbers.

%
%
\section{Asymptotics}
\label{hypergeom}

We here discuss the asymptotics of two hypergeometric 
functions that are often encountered in 
bound-state calculations of the 
kind~\cite{JePa1996,JeEtAl2003,Pa1993,JeSoMo1997} 
considered in the 
current brief review. The first of these functions is
\begin{equation}
\Phi_1(n, t) = {}_2F_1(1, -n t, 1- n t, \xi) = 
- n t \sum_{k=0}^{\infty} \frac{\xi^k}{k-nt}\,,
\end{equation}
which develops singularities at $t = k/n$. These correspond to 
lower-lying states and their equidistant spacing is a 
consequence of the energy parameterization discussed 
in Sec.~\ref{dimless}.  For given $n$, there are
typically $n-1$ lower-lying states accessible by dipole decay,
and thus $n-1$ singularities at $t = m/n$, for $m = 1,\dots n-1$,
as $t$ is in the range $(0,1)$. The argument $\xi$ of the 
hypergeometric typically reads
\begin{equation}
\xi = \left( \frac{1-t}{1+t} \right)^2\,.
\end{equation}
The small-$t$ asymptotics read
\begin{align}
\Phi_1(n, t) =& 
1 + n t \ln(4 t)
- \left[ 2 n + n^2 \zeta(2) \right] t^2 
\nonumber\\[2ex]
& + \left\{ n + 4 n^2 \left[  1- \ln (4 t)\right] - n^3 \zeta(3) \right\} t^3
\nonumber\\[2ex]
& + \left[ -\frac23\,n + 4 n^2 + 4 n^3 \zeta(2) - n^4 \zeta(4)\right] t^4 + 
{\cal O}\left[t^5 \, \ln(t) \right]\,.
\end{align}

The other function which is often encountered reads
\begin{align}
\Phi_2(n, \, t, \, \zeta) &= {}_2F_1(1, \, -nt, \, 1-nt, \, \zeta) 
= -nt \sum_{k=0}^{\infty} \frac{(-\zeta)^k}{k-nt} \,,
\end{align}
where
\begin{equation}
\zeta = \frac{1-t}{1+t}\,.
\end{equation}
Its asymptotics are given by
\begin{align}
\Phi_2(n, t) =& 
1 + n t \ln(2)
+ \left[ \frac12\,n^2 \zeta(2) - n \right] t^2 
\nonumber\\[2ex]
& + \left[ \frac{n}{2} - 2 n^2 \ln(2) + \frac34 n^3 \zeta(3) \right] t^3
\nonumber\\[2ex]
& + \left[ -\frac13\,n + n^2 - n^3 \, \zeta(2) + 
\frac78 n^4 \zeta(4)\right] t^4 + 
{\cal O}\left[t^5 \, \ln(t) \right]\,.
\end{align}
Numerical algorithms useful for different $t$-ranges are
discussed in Table 3. For a description
of the combined nonlinear-condensation 
transformation (CNCT), the reader is referred to 
Refs.~\cite{JeMoSoWe1999,AkSaJeBeSoMo2003}.

\begin{table}[ht]
\begin{center}
\begin{minipage}{10cm}
\begin{center}
\caption{Numerical algorithms used in the calculation of the 
functions $\Phi_1$ and $\Phi_2$.}
\begin{tabular}{ccc} 
\hline
\hline
& $0 < t < 0.05$ & $0.05 < t < 1$ \\ 
\hline
$\Phi_1(n, \, t, \, \xi)$   & CNCT & power series$+$recursion \\ 
$\Phi_2(n, \, t, \, \zeta)$ & $\delta$ transformation & 
power series$+$recursion \\ 
\hline
\hline
\end{tabular}
\end{center}
\end{minipage}
\end{center}
\end{table}

%
%
\section*{Acknowledgments}

The author acknowledges support from the Deutsche Forschungsgemeinschaft
(Heisenberg program).
Enlightening and insightful discussions with Krzysztof Pachucki
and Peter J.~Mohr are gratefully acknowledged.
The author thanks E.~O.~Le Bigot for help in the interpolation and
extrapolation of analytic results obtained in previous
calculations, to the region 
of higher nuclear charge numbers.
S. Jentschura is acknowledged for suggesting the 
possible derivation of exact results for the high-energy 
contributions to the Lamb shift,
as listed in Eq.~(12) of Ref.~\cite{JeEtAl2003},
and for carefully reading the manuscript.


\begin{thebibliography}{10}

\bibitem{Fe1948}
R.~P. Feynman, Phys. Rev. {\bf 74},  1430  (1948).

\bibitem{KrLa1949}
N.~M. Kroll and W.~E. Lamb, Phys. Rev. {\bf 75},  388  (1949).

\bibitem{FrWe1949}
J.~B. French and V.~F. Weisskopf, Phys. Rev. {\bf 75},  1240  (1949).

\bibitem{Fe1949}
R.~P. Feynman, Phys. Rev. {\bf 76},  769  (1949).

\bibitem{BeLiPi1982}
V.~B. Berestetskii, E.~M. Lifshitz, and L.~P. Pitaevskii, {\em Quantum
  Electrodynamics} (Pergamon Press, Oxford, UK, 1982).

\bibitem{CaLe1986}
W.~E. Caswell and G.~P. Lepage, Phys. Lett. B {\bf 167},  437  (1986).

\bibitem{KiNi1996}
T. Kinoshita and M. Nio, Phys. Rev. D {\bf 53},  4909  (1996).

\bibitem{NiKi1997}
M. Nio and T. Kinoshita, Phys. Rev. D {\bf 55},  7267  (1997).

\bibitem{PiSo1998}
A. Pineda and J. Soto, Phys. Rev. D {\bf 59},  016005  (1998).

\bibitem{NiEtAl2000}
M. Niering, R. Holzwarth, J. Reichert, P. Pokasov, T. Udem, M. Weitz, T.~W.
  H\"{a}nsch, P. Lemonde, G. Santarelli, M. Abgrall, P. Laurent, C. Salomon,
  and A. Clairon, Phys. Rev. Lett. {\bf 84},  5496  (2000).

\bibitem{UdEtAl1997}
T. Udem, A. Huber, B. Gross, J. Reichert, M. Prevedelli, M. Weitz, and T.~W.
  H\"{a}nsch, Phys. Rev. Lett. {\bf 79},  2646  (1997).

\bibitem{BeEtAl1997}
B. de~Beauvoir, F. Nez, L. Julien, B. Cagnac, F. Biraben, D. Touahri, L.
  Hilico, O. Acef, A. Clairon, and J.~J. Zondy, Phys. Rev. Lett. {\bf 78},  440
   (1997).

\bibitem{JePa1996}
U. Jentschura and K. Pachucki, Phys. Rev. A {\bf 54},  1853  (1996).

\bibitem{JeKoLBMoTa2005}
U.~D. Jentschura, S. Kotochigova, E.-O. Le~Bigot, P.~J. Mohr, and B.~N. Taylor,
  Precise calculation of hydrogenic energy levels using the method of least
  squares, Phys. Rev. Lett., in press (2005).

\bibitem{MoTa2005}
P.~J. Mohr and B.~N. Taylor, Rev. Mod. Phys. {\bf 77},  1  (2005).

\bibitem{Po1981}
A. Poquerusse, Phys. Lett. A {\bf 82},  232  (1981).

\bibitem{Je2003jpa}
U.~D. Jentschura, J. Phys. A {\bf 36},  L229  (2003).

\bibitem{JeEtAl2003}
U.~D. Jentschura, E.-O. Le~Bigot, P.~J. Mohr, P. Indelicato, and G. Soff, Phys.
  Rev. Lett. {\bf 90},  163001  (2003).

\bibitem{Wo1988}
S. Wolfram, {\em Mathematica-A System for Doing Mathematics by Computer}
  (Addison-Wesley, Reading, MA, 1988).

\bibitem{LBEtAl2003}
E.-O. Le~Bigot, U.~D. Jentschura, P.~J. Mohr, P. Indelicato, and G. Soff, Phys.
  Rev. A {\bf 68},  042101  (2003).

\bibitem{Pa1993}
K. Pachucki, Ann. Phys. (N.Y.) {\bf 226},  1  (1993).

\bibitem{Je2003dipl}
U.~D. Jentschura, Theory of the Lamb Shift in Hydrogenlike Systems, e-print
  hep-ph/0305065; based on an unpublished ``Master Thesis: The Lamb Shift in
  Hydrogenlike Systems'', [in German: ``Theorie der Lamb--Verschiebung in
  wasserstoffartigen Systemen''], Ludwig--Maximilians--University of Munich,
  Germany (1996).

\bibitem{ItZu1980}
C. Itzykson and J.~B. Zuber, {\em Quantum Field Theory} (McGraw-Hill, New York,
  NY, 1980).

\bibitem{JePa2002}
U.~D. Jentschura and K. Pachucki, J. Phys. A {\bf 35},  1927  (2002).

\bibitem{JeKePa2002}
U.~D. Jentschura, C.~H. Keitel, and K. Pachucki, Can. J. Phys. {\bf 80},  1213
  (2002).

\bibitem{JeEvKePa2002}
U.~D. Jentschura, J. Evers, C.~H. Keitel, and K. Pachucki, New J. Phys. {\bf
  4},  49  (2002).

\bibitem{Pa2004}
K. Pachucki, Phys. Rev. A {\bf 69},  052502  (2004).

\bibitem{JeSoMo1997}
U.~D. Jentschura, G. Soff, and P.~J. Mohr, Phys. Rev. A {\bf 56},  1739
  (1997).

\bibitem{SwDr1991a}
R.~A. Swainson and G.~W.~F. Drake, J. Phys. A {\bf 24},  79  (1991).

\bibitem{SwDr1991b}
R.~A. Swainson and G.~W.~F. Drake, J. Phys. A {\bf 24},  95  (1991).

\bibitem{SwDr1991c}
R.~A. Swainson and G.~W.~F. Drake, J. Phys. A {\bf 24},  1801  (1991).

\bibitem{SaOe1989}
S. Salomonson and P. \"{O}ster, Phys. Rev. A {\bf 40},  5559  (1989).

\bibitem{PaJe2003}
K. Pachucki and U.~D. Jentschura, Phys. Rev. Lett. {\bf 91},  113005  (2003).

\bibitem{Je2004b60}
U.~D. Jentschura, Phys. Rev. A {\bf 70},  052108  (2004).

\bibitem{Pa1999}
K. Pachucki, J. Phys. B {\bf 32},  137  (1999).

\bibitem{Pi2002pra}
A. Pineda, Phys. Rev. A {\bf 66},  062108  (2002).

\bibitem{Pi2002prd}
A. Pineda, Phys. Rev. D {\bf 66},  054022  (2002).

\bibitem{SaYe1990}
J. Sapirstein and D.~R. Yennie,  in {\em Quantum Electrodynamics}, Vol.~7 of
  {\em Advanced Series on Directions in High Energy Physics}, edited by T.
  Kinoshita (World Scientific, Singapore, 1990), pp.\ 560--672.

\bibitem{MoTa2000}
P.~J. Mohr and B.~N. Taylor, Rev. Mod. Phys. {\bf 72},  351  (2000).

\bibitem{Be1947}
H.~A. Bethe, Phys. Rev. {\bf 72},  339  (1947).

\bibitem{JeMoSo1999}
U.~D. Jentschura, P.~J. Mohr, and G. Soff, Phys. Rev. Lett. {\bf 82},  53
  (1999).

\bibitem{JeMoSo2001pra}
U.~D. Jentschura, P.~J. Mohr, and G. Soff, Phys. Rev. A {\bf 63},  042512
  (2001).

\bibitem{JeMo2004pra}
U.~D. Jentschura and P.~J. Mohr, Phys. Rev. A {\bf 69},  064103  (2004).

\bibitem{JeMoSoWe1999}
U.~D. Jentschura, P.~J. Mohr, G. Soff, and E.~J. Weniger, Comput. Phys. Commun.
  {\bf 116},  28  (1999).

\bibitem{AkSaJeBeSoMo2003}
S.~V. Aksenov, M.~A. Savageau, U.~D. Jentschura, J. Becher, G. Soff, and P.~J.
  Mohr, Comput. Phys. Commun. {\bf 150},  1  (2003).

\end{thebibliography}
\end{document}